\documentclass[a4paper,11pt]{article}
\usepackage{amsmath}
\usepackage{amssymb}
\usepackage{theorem}
\usepackage{multirow}
\usepackage{txfonts}
\usepackage{color}
\DeclareMathAlphabet{\bm}{OML}{cmm}{b}{it}
\newcommand{\textchange}[1]{#1}
\newcommand{\textchangesecond}[1]{#1}
\newcommand{\textchangethird}[1]{#1}
\newcommand{\textchangeforth}[1]{#1}
\theorembodyfont{\rmfamily}
\newtheorem{theorem}{Theorem}
\newtheorem{lemma}[theorem]{Lemma}

\newtheorem{corollary}[theorem]{Corollary}
\newtheorem{remark}[theorem]{Remark}
\newcommand{\qed}{\hfill$\square$}

\newcommand{\braket}[2]{\langle #1 | #2 \rangle} 
\newcommand{\ket}[1]{| #1 \rangle}
\newcommand{\bra}[1]{\langle #1 |}
\newcommand{\bol}[1]{\mathbf{#1}}
\newcommand{\rom}[1]{\mathrm{#1}}
\newcommand{\san}[1]{\mathsf{#1}}
\newcommand{\mymid}{:~}

\title{Optimal Axis Compensation in Quantum Key Distribution Protocols
over Unital Channels\thanks{A part of this paper will be presented at 2009
IEEE International Symposium on Information Theory. The results in
Section \ref{subsubsec:any-direction-solution} is additionally shown in 
this paper.}}
\author{Shun Watanabe\thanks{shun-wata@is.tokushima-u.ac.jp} \\
Department of Information Science and Intelligent Systems, \\
Tokushima University, \\
2-1, Minami-josanjima, Tokushima, 770-8506 Japan \\
Ryutaroh Matsumoto\thanks{ryutaroh@rmatsumoto.org}, 
and Tomohiko Uyematsu\thanks{uyematsu@ieee.org} \\
Department of Communications and Integrated Systems, \\
Tokyo Institute of Technology, \\
2-12-1, Oookayama, Meguro-ku, Tokyo, 152-8552, Japan 
}
\date{April 27, 2009}

\begin{document}
\maketitle

\begin{abstract}
\textchange{The axis compensation is a procedure in which
the sender and the receiver compensate the axes
of their transmitter and detector so that 
the bit sequence can be transmitted more reliably.}
We show the optimal axis compensations maximizing the key generation
 rate for unital channels. We consider the case in which only Bob is
allowed to compensate his axis, \textchange{and} 
the case in which both Alice and
Bob are allowed to compensate their axes. In the former case, we show
that we should utilize the mismatched measurement outcomes
in the channel estimation phase. In the latter case, we show that
we do not have to utilize the mismatched measurement outcomes
in the channel estimation phase.
\end{abstract}

\section{Introduction}

Quantum key distribution (QKD) has attracted great attention
as a technology to realize the information theoretically secure
key agreement. 
In this paper, we investigate the 
Bennett-Brassard 1984 (BB84) protocol \cite{bennett:84}
and the six-state protocol \cite{bruss:98}, and
\textchange{the QKD protocols indicate the BB84 protocol
and the six-state protocol.}

Typically in theoretical studies on 
the QKD protocols, the protocols roughly consist of three phases:
the bit transmission phase, the channel estimation phase,
and the postprocessing phase.
In the bit transmission phase, the legitimate sender, usually referred
to as Alice, sends a bit sequence to the legitimate receiver, usually
referred to as Bob, by encoding them into qubits. 
In the channel estimation phase, Alice and Bob estimate the channel and the amount
of information gained by an eavesdropper, usually referred to as
Eve. Finally in the postprocessing phase, Alice and Bob share
a secret key based on their bit sequences obtained in
the bit transmission phase.

On the other hand, in the practical QKD protocols, 
Alice and Bob conduct   
the {\em axis compensation} (before the bit transmission phase), in which 
Alice and Bob compensate the
axes of their transmitter and detector so that
the bit sequence can be transmitted more reliably
in the bit transmission phase.
This axis compensation is 
considered to be indispensable in the
QKD protocols, and it has been extensively studied from
the experimental point of view 
\cite{chen:07, franson:95, ma:06b, tao:06, xavier:08, zavriyev:05, trifonov:07}
(see also \cite{gisin:02}).
However, it has not been theoretically clarified how
Alice and Bob should compensate the axes in the axis compensation.

In this paper, we investigate the optimal axis compensation
in the sense that the key generation rate
is maximized, where 
the key generation rate is defined as the ratio between the length of
the shared \textchange{secret} key and that of the sequences 
initially possessed by Alice and Bob in the postprocessing phase.

We consider the following various settings.
In the channel estimation phase, we consider two kinds
of channel estimation: the accurate channel estimation
and the conventional channel estimation (see \cite{watanabe:08}).
In the accurate channel estimation, we use the 
mismatched measurement outcomes, which are bits
transmitted and received by different bases, in addition
to the matched measurement outcomes, which are
bits transmitted and received by the same bases,
to estimate the channel. 
On the other hand, in the conventional 
channel estimation, we discard the mismatched measurement
outcomes and only use the matched measurement outcomes.
The reason why we consider two kind of channel estimation 
is that the authors recently clarified that the key generation
rate is increased if we use the accurate channel estimation
instead of the conventional channel estimation \cite{watanabe:08}.
It is worthwhile to clarify whether we should use the accurate
channel estimation instead of the conventional channel estimation
when the QKD protocols involve the axis compensation.

In the postprocessing phase,
we employ the standard postprocessing. We do not use
the noisy preprocessing \cite{renner:05, kraus:05} 
nor the two-way classical communication \cite{gottesman:03, watanabe:07}.

In the axis compensation phase, we consider two kinds of
compensations:
\begin{enumerate}
\renewcommand{\theenumi}{\roman{enumi}}
\renewcommand{\labelenumi}{(\theenumi)}
\item 
\label{setting1}
({\em one-side compensation})
Only Bob is allowed to compensate his axis.

\item 
\label{setting2}
({\em two-side compensation})
Both Alice and Bob are allowed to compensate their
axes. 
\end{enumerate}

Furthermore in the BB84 protocol, we subdivide
each compensation into two kinds.
In the first kind, Bob (or both Alice and Bob) is allowed to
compensate his axis within the $\san{z}$-$\san{x}$
plane of the Bloch sphere. In the second kind, 
Bob (or both Alice and Bob) is allowed to compensate
his axis within any direction.
The reason why we consider these two kind of compensations
in the BB84 protocol is as follows. Since we only use the $\san{z}$-basis
and $\san{x}$-basis in the BB84 protocol, it is natural
to consider the axis compensation within the $\san{z}$-$\san{x}$ plane.
On the other hand, we might use the axis compensation within
any direction if we are allowed to enhance the device for
the compensation. Indeed, 
\textchange{several researchers} 
employ the compensation within
any direction in the literature 
\cite{chen:07, franson:95, ma:06b, tao:06, xavier:08, zavriyev:05, trifonov:07}.
Therefore, we also investigate the compensation within any direction.

The optimized key generation rates (of the standard 
postprocessing)
$F_1({\cal E})$, $\tilde{F}_1({\cal E})$, $F_2({\cal E})$,
$\tilde{F}_2({\cal E})$, $G_1({\cal E})$, $\tilde{G}_1({\cal E})$,
$G_2({\cal E})$, $\tilde{G}_2({\cal E})$, $J_1({\cal E})$,
$\tilde{J}_1({\cal E})$, $J_2({\cal E})$, $\tilde{J}_2({\cal E})$
for above described $12$ settings are summarized in
Table \ref{table:summary}. These quantities are
formally defined in Sections \ref{subsec:problem-six}
and \ref{subsec:problem-bb84} respectively.
\begin{table}[htbp]
\begin{center}
\label{table:summary}
\caption{Summary of the optimized key generation rates 
for various settings.}
\begin{tabular}{|c|c|c|c|c|} \hline
\multicolumn{3}{|c|}{\textchangethird{channel estimation}}  
  & accurate & conventional \\ \hline
\multirow{2}{*}{six-state} & \multicolumn{2}{c|}{one-side}  &
 {\scriptsize $F_1({\cal E})$} &
 {\scriptsize $\tilde{F}_1({\cal E})$} \\ \cline{2-5}
 & \multicolumn{2}{c|}{two-side}  & {\scriptsize $F_2({\cal E})$} & {\scriptsize $\tilde{F}_2({\cal E})$} \\
\hline
\multirow{4}{*}{BB84} & \multirow{2}{*}{$\san{z}$-$\san{x}$ plane} &
 one-side & {\scriptsize $G_1({\cal E})$} & {\scriptsize $\tilde{G}_1({\cal E})$} \\ \cline{3-5}
 & & two-side & {\scriptsize $G_2({\cal E})$} & {\scriptsize $\tilde{G}_2({\cal E})$} \\ \cline{2-5}
 & \multirow{2}{*}{any direction} & one-side & {\scriptsize $J_1({\cal E})$} &
 {\scriptsize $\tilde{J}_1({\cal E})$} \\ \cline{3-5} 
 & & two-side & {\scriptsize $J_2({\cal E})$} & {\scriptsize $\tilde{J}_2({\cal E})$} \\
\hline 
\end{tabular}
\end{center}
\end{table}

In this paper, we investigate the above described 
optimized key generation rates, and derive
closed-form expression of 
\textchangesecond{$F_1({\cal E})$, $F_2({\cal E})$,
$G_1({\cal E})$, $G_2({\cal E})$, $J_1({\cal E})$,
and $J_2({\cal E})$}
for unital channels.
\textchangethird{
Since QKD protocols can be implemented over many different media,
such as an optical fiber, free space, and an unknown medium,
we should conduct theoretical research with general quantum channels.
In this paper we deal with unital channels because we have closed-form
expressions of key generations rates of the QKD protocols.
The existence of such closed-form expressions enables us to identify
optimal compensation procedures. Without them identification is
difficult.}

By using the closed-form expressions of the optimized key
generation rates, we also derive the following relationships:
\begin{eqnarray*}
\begin{array}{ccccccc}
F_2({\cal E}) & = & \tilde{F}_2({\cal E}) &  = &  F_1({\cal E}) & \ge &
 \tilde{F}_1({\cal E}) \textchange{,} \\
G_2({\cal E}) & = & \tilde{G}_2({\cal E}) & = & G_1({\cal E}) & \ge &
 \tilde{G}_1({\cal E}) \textchange{,} \\
J_2({\cal E}) & = & \tilde{J}_2({\cal E}) & \ge & J_1({\cal E}) & \ge &
 \tilde{J}_1({\cal E})
\end{array}
\end{eqnarray*}
hold for any unital channel, and
\begin{eqnarray*}
F_1({\cal E}) &>& \tilde{F}_1({\cal E}) \textchange{,} \\
G_1({\cal E}) &>& \tilde{G}_1({\cal E}) \textchange{,} \\
J_1({\cal E}) &>& \tilde{J}_1({\cal E})
\end{eqnarray*}
hold for general cases of unital channels.

Our results provide the following important
insight. In the literatures 
\cite{chen:07, franson:95, ma:06b, tao:06, xavier:08, zavriyev:05, trifonov:07},
they employ the one-side compensation for the
axis compensation phase and the conventional channel
estimation for the channel estimation phase.
However, when we employ the one-side compensation, above
mentioned relationships imply that we should use
the accurate channel estimation. On the other hand,
when we employ the two-side compensation, above
mentioned relationships imply that we do not
have to use the accurate channel estimation. 

The rest of this paper is organized as follows:
In Section \ref{sec:problem}, we formally describe the
problem mentioned above for the six-state protocol and
the BB84 protocol. In Section \ref{sec:solution}, we
provide closed-form expressions of the optimized key
generation rates, and also clarify the relationships
among the optimized key generation rates for 
various settings. We state the conclusion in Section
\ref{sec:conclusion}.

\section{Problem Formulation}
\label{sec:problem}

In this section, we formally describe the problem
we investigate in this paper. 
Suppose that Alice and Bob are connected by a qubit
channel $\mathcal{E}_B$ from the set of all qubit density
operators to themselves. 
As is usual in QKD literatures, we assume that Eve can
access all the environment of channel $\mathcal{E}_B$; the channel
to the environment is denoted by $\mathcal{E}_E$.
In the rest of this paper, we omit the subscripts $B$ and $E$
if they are obvious from the context.

It should be noted that $\mathcal{E}$ can be any qubit channel
throughout the paper, unless we specify the channel to
be a Pauli channel or a unital channel.

\subsection{Stokes parameterization and Choi Operator}

For convenience, we introduce the Stokes parameterization
and the Choi operator for  the qubit channel. 
The qubit channel $\mathcal{E}$
can be described by the affine map parameterized
by $12$ real parameters \cite{fujiwara:98,fujiwara:99} as follows:
\begin{eqnarray}
\left[ \begin{array}{c}
\theta_{\san{z}} \\ \theta_{\san{x}} \\ \theta_{\san{y}}
\end{array} \right] 
\mapsto
\left[ \begin{array}{ccc}
R_{\san{zz}} & R_{\san{zx}} & R_{\san{zy}} \\
R_{\san{xz}} & R_{\san{xx}} & R_{\san{xy}} \\
R_{\san{yz}} & R_{\san{yx}} & R_{\san{yy}}
\end{array} \right]
\left[ \begin{array}{c}
\theta_{\san{z}} \\ \theta_{\san{x}} \\ \theta_{\san{y}}
\end{array} \right]
+ \left[ \begin{array}{c}
t_{\san{z}} \\ t_{\san{x}} \\ t_{\san{y}} 
\end{array} \right],
\label{eq-affine-map}
\end{eqnarray}
where $(\theta_\san{z}, \theta_\san{x}, \theta_\san{y})$
describes a vector in the Bloch sphere \cite{nielsen-chuang:00}.
The pair $(R,t)$ of the matrix and the vector
in Eq.~(\ref{eq-affine-map}) is called the Stokes parameterization
of the channel ${\cal E}$.
In the rest of this paper, we identify the channel ${\cal E}$
and its Stokes parameterization, and 
occasionally write ${\cal E} = (R, t)$.

For the channel $\mathcal{E}$ and each pair of bases
$(\san{a}, \san{b}) \in \{ \san{z}, \san{x}, \san{y} \}^2$,  
define the biases of the outputs as
\begin{eqnarray*}
Q_{\san{ab}0} &:=& 
\bra{0_{\san{b}}} \mathcal{E}_B( \ket{0_{\san{a}}} \bra{0_{\san{a}}}) \ket{0_{\san{b}}}
- \bra{1_{\san{b}}} \mathcal{E}_B( \ket{0_{\san{a}}} \bra{0_{\san{a}}}) \ket{1_{\san{b}}}, \\
Q_{\san{ab}1} &:=&
\bra{1_{\san{b}}} \mathcal{E}_B( \ket{1_{\san{a}}} \bra{1_{\san{a}}}) \ket{1_{\san{b}}}
- \bra{0_{\san{b}}} \mathcal{E}_B( \ket{1_{\san{a}}} \bra{1_{\san{a}}}) \ket{0_{\san{b}}},
\end{eqnarray*}
where $\ket{0_\san{a}}, \ket{1_\san{a}} $ are eigenstates of the Pauli 
operator $\sigma_\san{a}$ for $\san{a} \in \{\san{x},\san{y},\san{z}\}$ 
respectively.  
Then, a straight forward calculation shows the relations
\begin{eqnarray}
\label{eq-relation-between-bias-parameter}
R_{\san{ba}} = \frac{1}{2}(Q_{\san{ab}0} + Q_{\san{ab}1}),~~
t_{\san{b}} = \frac{1}{2}(Q_{\san{ab}0} - Q_{\san{ab}1}).
\end{eqnarray}

A unital channel is a channel that maps
the completely mixed state $I/2$ to itself.
For a unital channel, the vector part
$(t_\san{z}, t_\san{x}, t_\san{y})$ of the
Stokes parameterization is the zero vector. 
Furthermore, the channel is called a 
Pauli channel if the matrix part $R$ of
the Stokes parameterization is a diagonal
matrix in addition to that the vector part
is the zero vector.

We can also describe the qubit channel ${\cal E}$ 
by the Choi operator
\begin{eqnarray*}
\rho_{AB} := (\rom{id} \otimes {\cal E}_B)(\ket{\psi}\bra{\psi}),
\end{eqnarray*}
where $\ket{\psi} = \frac{\ket{00} + \ket{11}}{\sqrt{2}}$
is a maximally entangled state.
For the unital channel, the Choi operator satisfies
$\rom{Tr}_A[\rho_{AB}] = I/2$. Furthermore,
the channel is a Pauli channel if and only if
the Choi operator is a Bell diagonal operator, i.e.,
\begin{eqnarray*}
\rho_{AB} = \sum_{\san{a} \in \{\san{i},\san{z},\san{x},\san{y}\}}
q_\san{a} \ket{\psi_\san{a}}\bra{\psi_\san{a}}
\end{eqnarray*}
for Bell states 
$\ket{\psi_\san{a}} := (I \otimes \sigma_\san{a}) \ket{\psi}$
and the probability distribution $(q_\san{i},q_\san{z},q_\san{x},
q_\san{y})$ on $\{\san{i}, \san{z}, \san{x}, \san{y}\}$,
where 
$\sigma_\san{i}$ is the identity operator.
Throughout this paper, we omit the subscript $AB$ if it is
obvious from the context.

\subsection{Six-state protocol}
\label{subsec:problem-six}

As the preparation phase of the six-state protocol,
Alice and Bob conduct the following axis compensation
procedure.
Alice randomly sends bit $0$ or $1$ to Bob by modulating it
into a transmission basis that is randomly chosen from
the $\san{z}$-basis $\{ \ket{0_\san{z}}, \ket{1_\san{z}} \}$,
the $\san{x}$-basis $\{ \ket{0_\san{x}}, \ket{1_\san{x}} \}$, 
or the $\san{y}$-basis $\{ \ket{0_\san{y}}, \ket{1_\san{y}} \}$.
Then Bob randomly chooses one of measurement observables
$\sigma_\san{x}$, $\sigma_\san{y}$, and $\sigma_\san{z}$, and converts
a measurement result $+1$ or $-1$ into
a bit $0$ or $1$ respectively.
After a sufficient number of transmissions, Alice
and Bob publicly announce their transmission bases and
measurement observables. 
They also announce all of their bit sequences for
estimating channel $\mathcal{E}$.
Note that Alice and Bob do not discard mismatched measurement outcomes,
which are transmitted and received by different bases,
and they also use the mismatched measurement outcomes
to estimate the channel.
From Eq.~(\ref{eq-relation-between-bias-parameter}), we find
that Alice and Bob can estimate all parameters
$(R, t)$ of the channel in the six-state 
protocol \textchangesecond{\cite{chuang:97, poyatos:97}}.
\textchange{Note that the use of the mismatched measurement
outcomes for channel estimation is also known as the
process tomography.}

We consider two kinds of compensations:
\begin{enumerate}
\renewcommand{\theenumi}{\roman{enumi}}
\renewcommand{\labelenumi}{(\theenumi)}
\item Only Bob is allowed to compensate his axis, i.e.,
the channel after the compensation is
\begin{eqnarray}
\label{eq:compensated-channel-2}
\mathcal{E}_B^\prime = \mathcal{U}_B \circ \mathcal{E}_B,
\end{eqnarray}
where $\mathcal{U}_B$ is a unitary channel that represents
Bob's compensation.

\item Both Alice and Bob are allowed to compensate their
axes, i.e., the channel after the compensation is
\begin{eqnarray}
\label{eq:compensated-channel-1}
\mathcal{E}_B^\prime = \mathcal{U}_B \circ \mathcal{E}_B \circ \mathcal{U}_A,
\end{eqnarray}
where $\mathcal{U}_A$ and $\mathcal{U}_B$ are unitary channels
that represent Alice and Bob's compensations.
\end{enumerate}
Based on the estimate of the parameters $(R, t)$, Bob (or both Alice and
Bob) decides ${\cal U}_B$ (or ${\cal U}_A$ and ${\cal U}_B$), and
he compensates the channel. The choice of ${\cal U}_B$ (or ${\cal U}_A$
and ${\cal U}_A$) can be decided according to 
Theorem \ref{theorem:six} and Corollary \ref{corollary:six}
of Section \ref{sec:solution-six-state}.

\begin{remark}
Throughout this paper, the prime represents that it 
is after the compensation.
\end{remark}

After the compensation procedure, Alice and Bob conduct
the above bit transmission and reception procedure again.
This time, they only announce a part of their bit sequence
for estimating the channel $\mathcal{E}^\prime$, and 
they conduct the postprocessing to generate a secret key
from the remaining (unannounced) bit sequences.

Henceforth, we focus on the postprocessing procedure
for Alice's bit sequence $\bol{x} \in \mathbb{F}_2^n$
that is transmitted in $\san{z}$-basis and 
corresponding Bob's bit sequence
$\bol{y} \in \mathbb{F}_2^n$ that is received in
$\sigma_\san{z}$ measurement, where $\mathbb{F}_2$ 
is the finite field of order $2$.
We employ the standard postprocessing procedure that
consists of the information reconciliation procedure
and the privacy amplification procedure 
(e.g.~see \cite[Section 2]{watanabe:08}).
Note that we do not use the so-called noisy preprocessing
\cite{renner:05, kraus:05} nor the postprocessing with
two-way classical communication \cite{gottesman:03, watanabe:07}.

Let 
\begin{eqnarray*}
H_{\mathcal{E}^\prime}(X|E) := H(\rho^\prime_{XE}) - H(\rho^\prime_E)
\end{eqnarray*}
be the conditional von Neumann entropy with respect to
the density operator
\begin{eqnarray*}
\rho^\prime_{XE} := \sum_{x \in \mathbb{F}_2} \frac{1}{2}
  \ket{x_\san{z}}\bra{x_\san{z}} \otimes 
   \mathcal{E}_E^\prime(\ket{x_\san{z}}\bra{x_\san{z}})
\end{eqnarray*}
\textchangeforth{on the joint system ${\cal H}_X \otimes {\cal H}_E$ },
where $H(\rho)$ is the von Neumann entropy \cite{nielsen-chuang:00} for a 
density matrix $\rho$,  and
we take the base of the logarithm to be $2$ throughout
the paper.
\textchangethird{For the compensated channel ${\cal E}^\prime$,
we define the joint probability distribution}
\begin{eqnarray*}
P^\prime_{XY}(x,y) := \frac{1}{2}
  \bra{y_\san{z}} \mathcal{E}_B^\prime(\ket{x_\san{z}}\bra{x_\san{z}})
   \ket{y_\san{z}}
\end{eqnarray*}
\textchangeforth{of the joint random variable 
$(X,Y)$ on $\mathbb{F}_2 \times \mathbb{F}_2$}.
\textchangethird{Then, let}
\begin{eqnarray}
\label{eq:definition-of-conditional-entropy}
H_{\mathcal{E}^\prime}(X|Y) :=
  - \sum_{x, y \in \mathbb{F}_2} P_{XY}^\prime(x,y)
  \log P_{X|Y}^\prime(x|y)
\end{eqnarray}
be the conditional Shannon entropy of $X$ given $Y$.

In the six-state protocol, since Alice and Bob can estimate
the channel $\mathcal{E}^\prime$ exactly
if they use the accurate channel estimation, they can asymptotically share
a secure key if the length $\ell$ of the key satisfies
\begin{eqnarray*}
\frac{\ell}{n} < H_{\mathcal{E}^\prime}(X|E) 
  - H_{\mathcal{E}^\prime}(X|Y)
\end{eqnarray*}
(see \cite{renner:05b, watanabe:08}).
Therefore, we consider the following two
optimization problems:
\begin{enumerate}
\renewcommand{\theenumi}{\roman{enumi}}
\renewcommand{\labelenumi}{(\theenumi)}
\item Find a closed-form expression of
\begin{eqnarray}
\label{eq:optimization-F1}
F_1(\mathcal{E}) := 
\max_{\mathcal{U}_B} 
  [ H_{\mathcal{E}^\prime}(X|E) - H_{\mathcal{E}^\prime}(X|Y)],
\end{eqnarray}
and also find $\mathcal{U}_B$ \textchange{that} achieves
the maximum in Eq.~(\ref{eq:optimization-F1}).

\item Find a closed-form expression of 
\begin{eqnarray}
\label{eq:optimization-F2}
F_2(\mathcal{E}) := 
\max_{\mathcal{U}_A, \mathcal{U}_B} 
  [ H_{\mathcal{E}^\prime}(X|E) - H_{\mathcal{E}^\prime}(X|Y)],
\end{eqnarray}
and also find $(\mathcal{U}_A, \mathcal{U}_B)$ \textchange{that} achieve
the maximum in Eq.~(\ref{eq:optimization-F2}).
\end{enumerate}
Eqs.~(\ref{eq:optimization-F1}) and (\ref{eq:optimization-F2})
are the key generation rates optimized within the
one-side compensation and the two-side compensation respectively.

Next, we consider the case in which Alice and 
Bob use the conventional channel estimation (see \cite{watanabe:08}
for the detail of the conventional estimation).
From Eq.~(\ref{eq-relation-between-bias-parameter}),
we find that Alice and Bob can only estimate the parameters
$\gamma^\prime = (R_\san{zz}^\prime,R_\san{xx}^\prime,R_\san{yy}^\prime)$,
and they cannot estimate the parameters
$\kappa^\prime = (R_\san{zx}^\prime$, $R_\san{zy}^\prime$, $R_\san{xz}^\prime$, $R_\san{xy}^\prime$, $R_\san{yz}^\prime$, $R_\san{yx}^\prime$,
$t_\san{z}^\prime$, $t_\san{x}^\prime$, $t_\san{y}^\prime)$.
Since we have to consider the worst case with respect to the parameters
$\kappa^\prime$ that cannot be estimated, 
we consider the following two quantities:  
\begin{eqnarray}
\label{eq:optimization-tilde-F1}
\tilde{F}_1(\mathcal{E}) := 
\max_{\mathcal{U}_B} \min_{\tilde{\mathcal{E}} \in {\cal P}_s(\gamma^\prime)}
  [ H_{\tilde{\mathcal{E}}}(X|E) - H_{\tilde{\mathcal{E}}}(X|Y)]
\end{eqnarray}
and
\begin{eqnarray}
\label{eq:optimization-tilde-F2}
\tilde{F}_2(\mathcal{E}) := 
\max_{\mathcal{U}_A, \mathcal{U}_B} 
\min_{\tilde{\mathcal{E}} \in {\cal P}_s(\gamma^\prime)}
  [ H_{\tilde{\mathcal{E}}}(X|E) - H_{\tilde{\mathcal{E}}}(X|Y)],
\end{eqnarray}
where ${\cal P}_s(\gamma^\prime)$ is the set of all channel
for given $\gamma^\prime$, i.e.,
\begin{eqnarray*}
{\cal P}_s(\gamma^\prime) := \{ \tilde{\mathcal{E}} = (\tilde{\gamma},
 \tilde{\kappa}) \mymid  \tilde{\gamma} = \gamma^\prime \}.
\end{eqnarray*}
\textchange{Since the definition of $\tilde{F}_1({\cal E})$
and $\tilde{F}_2({\cal E})$ involve the minimization,
we have $F_1({\cal E}) \ge \tilde{F}_1({\cal E})$
and $F_2({\cal E}) \ge \tilde{F}_2({\cal E})$ \cite{watanabe:08}.}

\subsection{BB84 protocol}
\label{subsec:problem-bb84}

\subsubsection{Compensation within $\san{z}$-$\san{x}$ Plane}
\label{subsubsec:problem-bb84-1}

The BB84 protocol is almost the same as the 
six-state protocol. However in the BB84 protocol with
the axis compensation within the $\san{z}$-$\san{x}$ plane,
Alice uses only $\san{z}$ basis and $\san{x}$ basis
to transmit the bit sequence, and Bob uses only
observable $\sigma_\san{z}$ and $\sigma_\san{x}$ to
receive the bit sequence. Therefore, 
from Eq.~(\ref{eq-relation-between-bias-parameter}), we find that
Alice and Bob can only estimate the parameters
$\omega = (R_\san{zz},R_\san{zx}, R_\san{xz},
R_\san{xx}, t_\san{z}, t_\san{x})$, and that
they cannot estimate the parameters
$\tau = (R_\san{zy}, R_\san{xy}, R_\san{yz},
 R_\san{yx}, R_\san{yy}, t_\san{y})$.
We consider the following two kinds of compensations:
\begin{enumerate}
\renewcommand{\theenumi}{\roman{enumi}}
\renewcommand{\labelenumi}{(\theenumi)}
\item Only Bob is allowed to compensate 
his axis within the $\san{z}$--$\san{x}$ plane,
i.e., the channel after the compensation
is given by Eq.~(\ref{eq:compensated-channel-2}),
where $\mathcal{U}_B$ is a unitary channel that
rotate the Bloch sphere within the $\san{z}$--$\san{x}$ plane.

\item Both Alice and Bob are allowed to compensate
their axes within the $\san{z}$--$\san{x}$ plane,
i.e., the channel after the compensation is given by 
Eq.~(\ref{eq:compensated-channel-1}), where
$\mathcal{U}_A$ and $\mathcal{U}_B$ are unitary channels
that rotate the Bloch sphere within the $\san{z}$--$\san{x}$ plane.
\end{enumerate}
Based on the estimate of the parameters  $\omega$, 
Bob (or both Alice and Bob) decides ${\cal U}_B$
(or ${\cal U}_A$ and ${\cal U}_B$), and he compensates
the channel. The choice of ${\cal U}_B$ (or ${\cal U}_A$ and
${\cal U}_B$) can be decided according to Theorem \ref{theorem:bb84}
and Corollary \ref{corollary:bb84} of Section \ref{sec:solution-bb84-1}.

As in the six-state protocol, we also employ the 
standard postprocessing. 
We first consider the case in which Alice and Bob use
the accurate channel estimation. Note 
that Alice and Bob can only estimate the parameters
$\omega^\prime = (R_\san{zz}^\prime,R_\san{zx}^\prime, R_\san{xz}^\prime,
R_\san{xx}^\prime, t_\san{z}^\prime, t_\san{x}^\prime)$, and that
they cannot estimate the parameters
$\tau^\prime = (R_\san{zy}^\prime, R_\san{xy}^\prime, R_\san{yz}^\prime,
 R_\san{yx}^\prime, R_\san{yy}^\prime, t_\san{y}^\prime)$.

Let ${\cal P}_{ba}(\omega^\prime)$ be the set of all channels for
given $\omega^\prime$, i.e.,
\begin{eqnarray*}
{\cal P}_{ba}(\omega^\prime) := \{ \tilde{\mathcal{E}} = 
 (\tilde{\omega}, \tilde{\tau}) \mymid \tilde{\omega} = \omega^\prime \}.
\end{eqnarray*}
In the BB84 protocol, Alice and Bob can asymptotically share a secure key
if the length $\ell$ of the key satisfies 
\begin{eqnarray*}
\frac{\ell}{n} < \min_{\tilde{\mathcal{E}} \in {\cal P}_{ba}(\omega^\prime)}
 [ H_{\tilde{\mathcal{E}}}(X|E) - H_{\tilde{\mathcal{E}}}(X|Y) ]
\end{eqnarray*}
(see \cite{renner:05b, watanabe:08}).
Therefore, we consider the following two optimization
problems:
\begin{enumerate}
\renewcommand{\theenumi}{\roman{enumi}}
\renewcommand{\labelenumi}{(\theenumi)}
\item Find a closed-form expression of 
\begin{eqnarray}
G_1(\mathcal{E}) := 
\max_{\mathcal{U}_B} 
 \min_{\tilde{\mathcal{E}} \in {\cal P}_{ba}(\omega^\prime)} [
 H_{\tilde{\mathcal{E}}}(X|E) - H_{\tilde{\mathcal{E}}}(X|Y) ],
\label{eq:optimization-G1}
\end{eqnarray}
and also find $\mathcal{U}_B$ \textchange{that} achieves the maximum
in Eq.~(\ref{eq:optimization-G1}), where  
$\mathcal{U}_B$ is a unitary channel that rotates the Bloch sphere 
within the $\san{z}$--$\san{x}$ plane.

\item Find a closed-form expression of 
\begin{eqnarray}
G_2(\mathcal{E}) := 
\max_{\mathcal{U}_A, \mathcal{U}_B} 
\min_{\tilde{\mathcal{E}} \in {\cal P}_{ba}(\omega^\prime)}
 [ H_{\tilde{\mathcal{E}}}(X|E) - H_{\tilde{\mathcal{E}}}(X|Y) ],
\label{eq:optimization-G2}
\end{eqnarray}
and also find $(\mathcal{U}_A, \mathcal{U}_B)$ 
\textchange{that} achieve the maximum
in Eq.~(\ref{eq:optimization-G2}), where $\mathcal{U}_A$ and 
$\mathcal{U}_B$ are unitary channels that rotate the Bloch sphere 
within the $\san{z}$--$\san{x}$ plane.
\end{enumerate}

Next, we consider the case in which Alice and 
Bob use the conventional channel estimation.
From Eq.~(\ref{eq-relation-between-bias-parameter}),
we find that Alice and Bob can only estimate the parameters
$\mu^\prime = (R_\san{zz}^\prime,R_\san{xx}^\prime)$,
and they cannot estimate the parameters
$\nu^\prime = (R_\san{zx}^\prime$, $R_\san{zy}^\prime$, $R_\san{xz}^\prime$, $R_\san{xy}^\prime$, $R_\san{yz}^\prime$, $R_\san{yx}^\prime$, 
$R_\san{yy}^\prime$, $t^\prime_\san{z}$, 
$t_\san{x}^\prime$, $t_\san{y}^\prime)$.
Since we have to consider the worst case with respect to the
parameters $\nu^\prime$ that cannot be estimated, 
we consider the following two quantities: 
\begin{eqnarray}
\tilde{G}_1(\mathcal{E}) := 
\max_{\mathcal{U}_B} 
 \min_{\tilde{\mathcal{E}} \in {\cal P}_{bc}(\mu^\prime)} [
 H_{\tilde{\mathcal{E}}}(X|E) - H_{\tilde{\mathcal{E}}}(X|Y) ],
\label{eq:optimization-tilde-G1}
\end{eqnarray}
and 
\begin{eqnarray}
\tilde{G}_2(\mathcal{E}) := 
\max_{\mathcal{U}_A, \mathcal{U}_B} 
 \min_{\tilde{\mathcal{E}} \in {\cal P}_{bc}(\mu^\prime)}
 [ H_{\tilde{\mathcal{E}}}(X|E) - H_{\tilde{\mathcal{E}}}(X|Y) ],
\label{eq:optimization-tilde-G2}
\end{eqnarray}
where ${\cal P}_{bc}(\mu^\prime)$ is the set of all channel
for given $\mu^\prime$, i.e.,
\begin{eqnarray*}
{\cal P}_{bc}(\mu^\prime) := \{ \tilde{\mathcal{E}} = (\tilde{\mu},
 \tilde{\nu}) \mymid  \tilde{\mu} = \mu^\prime \}.
\end{eqnarray*}
\textchange{Since the range of the minimizations
in the definitions of $G_1({\cal E})$, $G_2({\cal E})$,
$\tilde{G}_1({\cal E})$, and $\tilde{G}_2({\cal E})$
satisfy 
${\cal P}_{ba}(\omega^\prime) \subset {\cal P}_{bc}(\mu^\prime)$,
we have $G_1({\cal E}) \ge \tilde{G}_1({\cal E})$
and $G_2({\cal E}) \ge \tilde{G}_2({\cal E})$ \cite{watanabe:08}.}

\subsubsection{Compensation within Any Direction}
\label{subsubsec:problem-bb84-2}

In this section, we consider the BB84 protocol
with the axis compensation within any direction.
We consider this problem because \textchange{several researchers} employ the
compensation within any direction in the literatures
\cite{chen:07, franson:95, ma:06b, tao:06, xavier:08, zavriyev:05, trifonov:07}.

When we employ the one-side compensation, 
\textchange{Alice randomly sends $0$ or $1$ to Bob
by modulating it into a transmission basis that
is randomly chosen from the $\san{z}$-basis
or the $\san{x}$-basis. Then Bob measures received
qubits by randomly using observables $\sigma_\san{z}$,
$\sigma_\san{x}$ or $\sigma_\san{y}$. Note that
Bob can use $\sigma_\san{y}$ in addition to
$\sigma_\san{z}$ and $\sigma_\san{x}$ because
he is allowed to rotate the axis of the receiver
in the axis compensation phase.}
In this case, from Eq.~(\ref{eq-relation-between-bias-parameter}),
we find that Alice and Bob can estimate the parameters
$(R_\san{zz}, R_\san{xz},R_\san{yz}, R_\san{zx}, R_\san{xx}, R_\san{yx},
t_\san{z}, t_\san{x}, t_\san{y})$, and they cannot
estimate the parameters
$(R_\san{zy}, R_\san{xy}, R_\san{yy})$.
\textchange{Since Bob can use $\sigma_\san{y}$, Alice
and Bob can estimate $(R_\san{yz},R_\san{yx}, t_\san{y})$
in addition to $(R_\san{zz},R_\san{xz}, R_\san{zx}, R_\san{xx}, t_\san{z},t_\san{x})$, which can be estimated in the compensation scheme of
Section \ref{subsubsec:problem-bb84-1}.}
Based on the estimate of the parameters
$(R_\san{zz}, R_\san{xz},R_\san{yz}, R_\san{zx}, R_\san{xx}, R_\san{yx},
t_\san{z}, t_\san{x}, t_\san{y})$, Bob decide ${\cal U}_B$ and
compensate the channel. The choice of ${\cal U}_B$ can be decided
according to Theorem \ref{theorem:bb84-3} of Section
\ref{subsubsec:any-direction-solution}.

On the other hand, when we employ the two-side
compensation, we allow both Alice and Bob to
use $\san{z}$-basis, $\san{x}$-basis, and $\san{y}$-basis
in the axis compensation phase.
In this case, from Eq.~(\ref{eq-relation-between-bias-parameter}),
we find that Alice and Bob can estimate all of the parameters
$(R, t)$. Based on the estimate of the parameters $(R, t)$,
Alice and Bob decide ${\cal U}_A$ and ${\cal U}_B$, and 
they compensate the channel. The choice of ${\cal U}_A$
and ${\cal U}_B$ can be decided according to 
Theorem \ref{theorem:bb84-2} of Section \ref{subsubsec:any-direction-solution}.

In the bit transmission phase (after the axis compensation phase),
we allow Alice and Bob to use only $\san{z}$-basis and 
$\san{x}$-basis.
The channel estimation phase and the postprocessing
phase are exactly the same as in Section \ref{subsubsec:problem-bb84-1}.
\textchange{Note that Alice and Bob can estimate
\textchangethird{$(R_\san{zz}^\prime,R_\san{xz}^\prime, R_\san{zx}^\prime, R_\san{xx}^\prime, t_\san{z}^\prime,t_\san{x}^\prime)$},
but they cannot estimate the other parameters,
because we do not allow 
neither Alice nor Bob to use $\san{y}$-basis
in the bit transmission phase.}
Therefore, we consider the following two optimization problems:
\begin{enumerate}
\renewcommand{\theenumi}{\roman{enumi}}
\renewcommand{\labelenumi}{(\theenumi)}
\item Find a closed-form expression of
\begin{eqnarray}
J_1(\mathcal{E}) := 
\max_{\mathcal{U}_B} 
 \min_{\tilde{\mathcal{E}} \in {\cal P}_{ba}(\omega^\prime)} [
 H_{\tilde{\mathcal{E}}}(X|E) - H_{\tilde{\mathcal{E}}}(X|Y) ],
\label{eq:optimization-J1}
\end{eqnarray}
and also find $\mathcal{U}_B$ \textchange{that} achieve the maximum
in Eq.~(\ref{eq:optimization-J1}), where  
$\mathcal{U}_B$ is any unitary channel.

\item Find a closed-form expression of 
\begin{eqnarray}
J_2(\mathcal{E}) := 
\max_{\mathcal{U}_A, \mathcal{U}_B} 
\min_{\tilde{\mathcal{E}} \in {\cal P}_{ba}(\omega^\prime)}
 [ H_{\tilde{\mathcal{E}}}(X|E) - H_{\tilde{\mathcal{E}}}(X|Y) ],
\label{eq:optimization-J2}
\end{eqnarray}
and also find $(\mathcal{U}_A, \mathcal{U}_B)$ \textchange{that} 
achieve the maximum
in Eq.~(\ref{eq:optimization-J2}), where $\mathcal{U}_A$ and 
$\mathcal{U}_B$ are any unitary channels.
\end{enumerate}

We also treat the case in which Alice and Bob use the
conventional channel estimation. In this case, we consider the
following two quantities:
\begin{eqnarray}
\tilde{J}_1(\mathcal{E}) := 
\max_{\mathcal{U}_B} 
 \min_{\tilde{\mathcal{E}} \in {\cal P}_{bc}(\mu^\prime)} [
 H_{\tilde{\mathcal{E}}}(X|E) - H_{\tilde{\mathcal{E}}}(X|Y) ],
\label{eq:optimization-tilde-J1}
\end{eqnarray}
and
\begin{eqnarray}
\tilde{J}_2(\mathcal{E}) := 
\max_{\mathcal{U}_A, \mathcal{U}_B} 
\min_{\tilde{\mathcal{E}} \in {\cal P}_{bc}(\mu^\prime)}
 [ H_{\tilde{\mathcal{E}}}(X|E) - H_{\tilde{\mathcal{E}}}(X|Y) ],
\label{eq:optimization-tilde-J2}
\end{eqnarray}
where ${\cal U}_A$ and ${\cal U}_B$ are any unitary channels.
\textchange{Since the range of the minimizations
in the definitions of $J_1({\cal E})$, $J_2({\cal E})$,
$\tilde{J}_1({\cal E})$, and $\tilde{J}_2({\cal E})$
satisfy 
${\cal P}_{ba}(\omega^\prime) \subset {\cal P}_{bc}(\mu^\prime)$,
we have $J_1({\cal E}) \ge \tilde{J}_1({\cal E})$
and $J_2({\cal E}) \ge \tilde{J}_2({\cal E})$ \cite{watanabe:08}.}

\section{Optimal Compensation for Unital Channels}
\label{sec:solution}

In this section, we solve the problems 
formulated in Sections \ref{subsec:problem-six}, 
\ref{subsubsec:problem-bb84-1}, and \ref{subsubsec:problem-bb84-2} 
respectively for unital channels.

\subsection{Six-state protocol}
\label{sec:solution-six-state}

For any channel $\mathcal{E} = (R,t)$, by the 
singular value decomposition, we can decompose\footnote{The decomposition is not unique 
because we can change the order of 
$(e_\san{z}, e_\san{x}, e_\san{y})$ or
the sign of them by adjusting the rotation matrices $A$ and $B$.
However, the result in this paper does not depends on a choice of the decomposition.}
the matrix $R$ as
\begin{eqnarray}
R &=& B ~\mbox{diag}[ e_\san{z}, e_\san{x}, e_\san{y} ] ~A \nonumber \\
&=& \left[ \begin{array}{c}
  \bra{B_\san{z}} \\ \bra{B_\san{x}} \\ \bra{B_\san{y}}
  \end{array} \right]
 \left[ \begin{array}{ccc} 
  e_\san{z} & 0 & 0 \\
  0 & e_\san{x} & 0 \\
  0 & 0 & e_\san{y}
  \end{array} \right] 
 \left[ \begin{array}{ccc}
 \\ \ket{A_\san{z}} & \ket{A_\san{x}} & \ket{A_\san{y}} \\
\\ \end{array} \right] \nonumber 
\\
&=& \left[ \begin{array}{ccc}
  \braket{B_\san{z}}{\tilde{A}_\san{z}} &
  \braket{B_\san{z}}{\tilde{A}_\san{x}} &
  \braket{B_\san{z}}{\tilde{A}_\san{y}} \\
  \braket{B_\san{x}}{\tilde{A}_\san{z}} &
  \braket{B_\san{x}}{\tilde{A}_\san{x}} &
  \braket{B_\san{x}}{\tilde{A}_\san{y}} \\
  \braket{B_\san{y}}{\tilde{A}_\san{z}} & 
  \braket{B_\san{y}}{\tilde{A}_\san{x}} &
  \braket{B_\san{y}}{\tilde{A}_\san{y}}
 \end{array} \right],
\label{eq:singular-value-decomposition}
\end{eqnarray}
where $A$ and $B$ are the rotation 
matrices\footnote{The rotation matrix is the real
orthogonal matrix with determinant $1$.}, 
$|e_\san{z}|$, $|e_\san{x}|$, and $|e_\san{y}|$ are the
singular values of $R$,
and we set 
$\bra{\tilde{A}_\san{z}} = (e_\san{z} A_\san{zz}, e_\san{x} A_\san{zx}, e_\san{y} A_\san{zy})$,
$\bra{\tilde{A}_\san{x}} = (e_\san{z} A_\san{xz}, e_\san{x} A_\san{xx}, e_\san{y} A_\san{xy})$, and 
$\bra{\tilde{A}_\san{y}} = (e_\san{z} A_\san{yz}, e_\san{x} A_\san{yx},
e_\san{y} A_\san{yy})$.

\textchange{Henceforth, we identify
Alice's compensation ${\cal U}_A$  
and Bob's compensation ${\cal U}_B$ with
the $3\times 3$ rotation matrices $O_A$ and $O_B$.
Then, the matrix part of the Stokes parameterization
of the compensated channel ${\cal E}^\prime = (R^\prime,t^\prime)$
is given by $R^\prime = O_B R O_A$.}

The following theorem gives a closed-form expression of the key generation
rate optimized by the two-side compensation.
\begin{theorem}
\label{theorem:six}
Suppose that ${\cal E}$ is a unital channel.
\textchange{ Let $O_A^* = A^{-1}$ and $O_B^* = B^{-1}$,}
\textchangesecond{and let ${\cal U}_A^*$ and 
${\cal U}_B^*$ be the unitary channels corresponding
to $O_A^*$ and $O_B^*$ respectively.}
\textchange{Then, the compensated channel 
${\cal E}^* = {\cal U}_B^* \circ {\cal E} \circ {\cal U}_A^*$ is
the Pauli channel such that
the matrix part of the Stokes parameterization
is given by $R^* = \mbox{diag}[e_\san{z}, e_\san{x}, e_\san{y}]$,
and ${\cal E}^*$ satisfies }
\begin{eqnarray}
\label{eq:statement-six-1}
F_2(\mathcal{E}) &=&  \max_{\mathcal{U}_A, \mathcal{U}_B} \left[
 H_{\mathcal{E}^\prime}(X|E) - H_{\mathcal{E}^\prime}(X|Y) \right]  \\
 \label{eq:statement-six-2}
&=& H_{\mathcal{E}^*}(X|E) - H_{\mathcal{E}^*}(X|Y) \\
 \label{eq:statement-six-3} 
&=& 1 - H[q_\san{i}, q_\san{z}, q_\san{x}, q_\san{y}],
\end{eqnarray}
where $H[q_\san{i}, q_\san{z}, q_\san{x}, q_\san{y}]$ is the 
Shannon entropy \cite{cover} of the distribution 
\begin{eqnarray}
\label{eq:pauli-1}
q_\san{i} &=& \frac{1 + e_\san{z} + e_\san{x} + e_\san{y}}{4}, \\
\label{eq:pauli-2}
q_\san{z} &=& \frac{1 + e_\san{z} - e_\san{x} - e_\san{y}}{4}, \\
\label{eq:pauli-3}
q_\san{x} &=& \frac{1 - e_\san{z} + e_\san{x} - e_\san{y}}{4}, \\
\label{eq:pauli-4}
q_\san{y} &=& \frac{1 - e_\san{z} - e_\san{x} + e_\san{y}}{4}.
\end{eqnarray}
Furthermore, the maximum in Eq.~(\ref{eq:statement-six-1})
is achieved without any compensation, i.e.,
\begin{eqnarray*}
H_{\mathcal{E}}(X|E) - H_{\mathcal{E}}(X|Y) 
 = H_{\mathcal{E}^*}(X|E) - H_{\mathcal{E}^*}(X|Y)
\end{eqnarray*}
if and only if the vectors $\ket{\tilde{A}_\san{z}}$ and
$\ket{B_\san{z}}$ are scalar multiple of each other.
\end{theorem}
The first statement implies that an optimal compensation
procedure is to compensate the channel to a Pauli channel.
The second statement implies that $({\cal U}_A, {\cal U}_B)$
\textchange{achieving} the maximum are not unique.

The following corollary gives a closed-form expression of
the key generation rate optimized by the one-side compensation.
\begin{corollary}
\label{corollary:six}
Suppose that ${\cal E}$ is a unital channel.
\textchange{Let}
\begin{eqnarray*}
\textchange{
O_B^* = \left[ \begin{array}{c}
\bra{O_{B,\san{z}}^*} \\
\bra{O_{B,\san{x}}^*} \\
\bra{O_{B,\san{y}}^*}
\end{array} \right]
}
\end{eqnarray*}
\textchange{be a rotation matrix such that
$\bra{O_{B,\san{z}}^*}$ is a scalar multiple of
$(R_\san{zz}, R_\san{xz}, R_\san{yz})$,
where $\bra{O_{B,\san{x}^*}}$ and $\bra{O_{B,\san{y}}^*}$
can be arbitrary as long as they constitute
a rotation matrix,} 
\textchangesecond{and let ${\cal U}_B^*$ the unitary channel
corresponding to $O_B^*$.}
\textchange{Then, the compensated channel
${\cal E}^* = {\cal U}_B^* \circ {\cal E}$ satisfies}
\begin{eqnarray*}
\textchange{ F_1(\mathcal{E}) }
&\textchange{=}& 
\textchange{ H_{{\cal E}^*}(X|E) - H_{{\cal E}^*}(X|Y) } \\
&\textchange{=}& \textchange{ F_2({\cal E}). }
\label{eq:corollary-six}
\end{eqnarray*} 
\textchangesecond{\qed}
\end{corollary}
\textchange{ Note that Corollary \ref{corollary:six} follows from the second
statement of Theorem \ref{theorem:six}.}

Surprisingly, we do not lose any optimality even if
we only allow Bob to compensate his axis (one-side compensation).
This fact is useful to simplify the implementation of the 
optimal compensation procedure.

Since  $H_{\cal E}(X|Y) = h((1 + R_\san{zz})/2)$ for any
unital channel and $R_\san{zz} = \braket{B_\san{z}}{\tilde{A}_\san{z}}$,  
we find that \textchange{an} optimal
one-side compensation procedure is to compensate
the channel so that Bob can detect Alice's transmitted
state most reliably, i.e.,
$H_{{\cal E}^\prime}(X|Y)$ is minimized,
where $h(\cdot)$ is the binary entropy function.
Note that the fact that $\ket{\tilde{A}_\san{z}}$ and $\ket{B_\san{z}^\prime}$
is scalar multiple of each other does not 
necessarily mean the compensated channel $\mathcal{E}^\prime$
is a Pauli channel. 

\noindent{\em Proof of Theorem \ref{theorem:six})~}
The equality between Eqs.~(\ref{eq:statement-six-2}) 
and (\ref{eq:statement-six-3})
is well known (e.g.~see \cite{renner:05} or
\cite[Eq.~(20)]{watanabe:08}).
\textchange{Since Eq.~(\ref{eq:statement-six-1}) 
is obviously larger than or equals to
Eq.~(\ref{eq:statement-six-2}),}
it \textchangethird{suffices} to show that Eq.~(\ref{eq:statement-six-1}) is smaller than
or equals to Eq.~(\ref{eq:statement-six-3}) for any ${\cal U}_A$ and
${\cal U}_B$. 
For any fixed ${\cal U}_A$ and ${\cal U}_B$, 
by using 
\cite[Eq.~(20)]{watanabe:08} and the discussions right before
it, Eq.~(\ref{eq:statement-six-1}) can be rewritten as
\begin{eqnarray*}
1 - H[q_\san{i},q_\san{z}, q_\san{x}, q_\san{y}]
 + h\left( \frac{1 + \| \ket{\tilde{A}^\prime_\san{z}} \| }{2} \right)
 - h\left( \frac{1 + \braket{B^\prime_\san{z}}{\tilde{A}^\prime_\san{z}}}{2} \right).
\end{eqnarray*}
From the form of $h(\cdot)$, 
Cauchy's inequality 
$|\braket{B^\prime_\san{z}}{\tilde{A}^\prime_\san{z}}| \le \| \ket{\tilde{A}^\prime_\san{z}} \|$ implies that Eq.~(\ref{eq:statement-six-1}) is smaller 
than or equals to Eq.~(\ref{eq:statement-six-3}). The equality holds
if and only if the vectors $\ket{\tilde{A}^\prime_\san{z}}$ and
$\ket{B^\prime_\san{z}}$ are scalar multiple of each other,
which is exactly the second statement of the theorem. 
\qed

Next, we consider the case in which Alice and Bob use the
conventional channel estimation. The following theorem
states that the optimized key generation rate with 
the accurate channel estimation \textchangethird{coincides} with that
with the conventional channel estimation if we use
the two-side compensation.
The following theorem also gives the
necessary and sufficient condition such that the
optimized key generation rates with the accurate
channel estimation and the conventional channel estimation
coincide when we use the one-side compensation.
\begin{theorem}
\label{theorem:six-2}
Suppose that ${\cal E}$ is a unital channel. Then, 
we have
\begin{eqnarray*}
F_2({\cal E}) = \tilde{F}_2({\cal E}),
\end{eqnarray*}
\textchangesecond{where $\tilde{F}_2({\cal E})$ is achieved by
$O_A^*$ and $O_B^*$ specified in Theorem \ref{theorem:six}.}
Furthermore, we have
\begin{eqnarray*}
F_1({\cal E}) = \tilde{F}_1({\cal E})
\end{eqnarray*}
if and only if
$\ket{\tilde{A}_\san{z}}$, $\ket{\tilde{A}_\san{x}}$, 
and $\ket{\tilde{A}_\san{y}}$ are orthogonal \textchange{to} each other.
\textchangesecond{If this condition is satisfied, then $\tilde{F}_1({\cal E})$
is achieved by $O_B^*$ such that
$\bra{O_{B,\san{z}}^*}$ and $\bra{O_{B,\san{x}}^*}$
and $\bra{O_{B,\san{y}}^*}$
are scalar multiple of 
$(R_\san{zz}, R_\san{xz}, R_\san{yz})$,
$(R_\san{zx}, R_\san{xx}, R_\san{yx})$,
and $(R_\san{zy}, R_\san{xy}, R_\san{yy})$
respectively.}
\end{theorem}
\begin{corollary}
Suppose that ${\cal E}$ is a unital channel. Then, 
we have
\begin{eqnarray*}
\tilde{F}_1({\cal E}) = \tilde{F}_2({\cal E})
\end{eqnarray*}
if and only if 
$\ket{\tilde{A}_\san{z}}$, $\ket{\tilde{A}_\san{x}}$, 
and $\ket{\tilde{A}_\san{y}}$ are orthogonal \textchange{to} each other.
\textchangesecond{\qed}
\end{corollary}
\noindent{\em Proof of Theorem \ref{theorem:six-2})~}
Let ${\cal E}^*$ be the Pauli channel defined in
Theorem \ref{theorem:six}. Then, we have
\begin{eqnarray*}
F_2({\cal E}) &\ge& 
\tilde{F}_2({\cal E}) \\
&\ge& \min_{\tilde{{\cal E}} \in {\cal P}_s(\gamma^*)}
[ H_{\tilde{{\cal E}}}(X|E) - H_{\tilde{{\cal E}}}(X|Y) ] \\
&=& 1 - H[q_\san{i}, q_\san{z}, q_\san{x}, q_\san{y} ] \\
&=& F_2({\cal E}),
\end{eqnarray*}
which implies the first statement of the theorem.

To prove the ``if'' part of the second statement, assume that
$\ket{\tilde{A}_\san{z}}$, $\ket{\tilde{A}_\san{x}}$, 
and $\ket{\tilde{A}_\san{y}}$ are orthogonal \textchange{to} each other.
\textchangesecond{Then, we can take a rotation matrix $O_B^*$ so that
$\bra{O_{B,\san{z}}^*}$ and $\bra{O_{B,\san{x}}^*}$
and $\bra{O_{B,\san{y}}^*}$
are scalar multiple of 
$(R_\san{zz}, R_\san{xz}, R_\san{yz})$,
$(R_\san{zx}, R_\san{xx}, R_\san{yx})$,
and $(R_\san{zy}, R_\san{xy}, R_\san{yy})$
respectively,
and we have
$R^\prime = O_B^* R = \mbox{diag}[e_\san{z},e_\san{x},e_\san{y}]$.}
Thus, we have
\begin{eqnarray*}
F_1({\cal E}) \ge \tilde{F}_1({\cal E}) \ge 1 - H[q_\san{i}, q_\san{z}, q_\san{x}, q_\san{y} ]
= F_1({\cal E}).
\end{eqnarray*}

Next, we show the ``only if'' part of the second statement.
Suppose that at least one pair of
$\ket{\tilde{A}_\san{z}}$, $\ket{\tilde{A}_\san{x}}$,
and $\ket{\tilde{A}_\san{y}}$ is not orthogonal \textchange{to} each other.
Then, for arbitrarily fixed ${\cal U}_B$, the compensated
channel ${\cal E}^\prime$ is not a Pauli channel, i.e.,
the Choi operator $\rho^\prime$ is not a Bell diagonal 
state. Let 
$\rho^\prime_\san{a} := (\bar{\sigma}_\san{a} \otimes \sigma_\san{a}) \rho^\prime (\bar{\sigma}_\san{a} \otimes \sigma_\san{a})$ for
$\san{a} \in \{ \san{i}, \san{z}, \san{x}, \san{y} \}$, 
where $\bar{\sigma}_\san{a}$ is the complex conjugate of $\sigma_\san{a}$.
Since $\rho^\prime$ is not Bell diagonal state, at least one of
$\rho^\prime_\san{z}$, $\rho^\prime_\san{x}$, and $\rho^\prime_\san{y}$
is different from $\rho_\san{i}^\prime$. Let 
\begin{eqnarray*}
\rho^{tw} := \sum_{\san{a} \in \{ \san{i}, \san{z}, \san{x}, \san{y} \}}
\frac{1}{4} \rho_\san{a}^\prime 
\end{eqnarray*}
be the partially twirled state \cite{bennett:96b}. Then, since the
von Neumann entropy is a strict concave function \cite{nielsen-chuang:00},
we have
\begin{eqnarray*}
\tilde{F}_1({\cal E}) &=&
\max_{{\cal U}_B}[ 1 - H(\rho^{tw})] \\
&<& \max_{{\cal U}_B}[ 1 - \sum_{\san{a} \in \{ \san{i}, \san{z}, \san{x}, 
\san{y} \}}
\frac{1}{4} H(\rho_\san{a}^\prime) ] \\
&=& \max_{{\cal U}_B}[ 1 - H(\rho^\prime)] \\
&=& F_1({\cal E}).
\end{eqnarray*}
\qed

\subsection{BB84 protocol}

\subsubsection{Compensation within $\san{z}$-$\san{x}$ Plane}
\label{sec:solution-bb84-1}

For any channel ${\cal E} = (R,t)$,
by the singular value decomposition, we can decompose
the left upper $2 \times 2$ sub-matrix $S$ of the matrix $R$ as
\begin{eqnarray*}
S &=& V~\mbox{diag}[d_\san{z}, d_\san{x}]~U \\
 &=& \left[ \begin{array}{c} 
  \bra{V_\san{z}} \\ \bra{V_\san{x}} \end{array} \right]
 \left[ \begin{array}{cc} d_\san{z} & 0 \\ 0 & d_\san{x} \end{array} \right]
  \left[ \begin{array}{cc} \ket{U_\san{z}} & \ket{U_\san{x}} \end{array} 
   \right] \\
 &=& \left[ \begin{array}{cc}
  \braket{V_\san{z}}{\tilde{U}_\san{z}} & \braket{V_\san{z}}{\tilde{U}_\san{x}} \\
  \braket{V_\san{x}}{\tilde{U}_\san{z}} & \braket{V_\san{x}}{\tilde{U}_\san{x}}
 \end{array} \right],
\end{eqnarray*}
where $U$ and $V$ are the rotation matrices, 
$|d_\san{z}|$ and $|d_\san{x}|$ are the singular 
values of $S$, and we set
$\bra{\tilde{U}_\san{z}} = (d_\san{z} U_\san{zz}, d_\san{x} U_\san{zx})$
and
$\bra{\tilde{U}_\san{x}} = (d_\san{z} U_\san{xz}, d_\san{x}
U_\san{xx})$.

\textchange{Henceforth, we identify
Alice's compensation ${\cal U}_A$  
and Bob's compensation ${\cal U}_B$ with
the $2 \times 2$ rotation matrices $Q_A$ and $Q_B$,
because their compensation are restricted within the
$\san{z}$-$\san{x}$ plane.
\textchangesecond{Note that} the left upper $2 \times 2$ sub-matrix $S^\prime$
of the matrix $R^\prime$ of the compensated channel is
given by $S^\prime = Q_B S Q_A$.}

The following lemma provides a closed-form expression
of the key generation rate with the accurate channel
estimation for unital channels, and it
will be
used several times in the rest of this paper.
\begin{lemma}
\label{lemma:unital-formula}
For any unital channel ${\cal E} = (\omega, \tau)$,
we have
\begin{eqnarray*}
\lefteqn{ \min_{\tilde{{\cal E}} \in {\cal P}_{ba}(\omega)} [
H_{\tilde{{\cal E}}}(X|E) - H_{\tilde{{\cal E}}}(X|Y)
] } \\
&=& 1 - h\left( \frac{1 + d_\san{z}}{2} \right)
 - h\left(\frac{1 + d_\san{x}}{2} \right)
+ h\left(\frac{1 + \sqrt{R_\san{zz}^2 + R_\san{xz}^2}}{2}\right)
 - h\left(\frac{1 + R_\san{zz}}{2}\right).
\end{eqnarray*}
\end{lemma}
\noindent{\em Proof of Lemma \ref{lemma:unital-formula})~}
This lemma follows from \cite[Proposition 2]{watanabe:08} and 
the fact $H_{{\cal E}}(X|Y) = h((1 + R_\san{zz})/2)$ for
any unital channel. \qed

The following theorem gives a closed-form expression of the
key generation rate optimized by the two-side compensation.
\begin{theorem}
\label{theorem:bb84}
Suppose that ${\cal E}$ is a unital channel.
\textchange{ Let $Q_A^* = U^{-1}$ and $Q_B^* = V^{-1}$,
\textchangesecond{and let ${\cal U}_A^*$ and 
${\cal U}_B^*$ be the unitary channels corresponding to
$Q_A^*$ and $Q_B^*$ respectively.}
Then, the compensated channel 
${\cal U}_B^* \circ {\cal E} \circ {\cal U}_A^* =: {\cal E}^* = (\omega^*,\tau^*)$
satisfies }
\begin{eqnarray}
\label{eq:theorem-bb84-1}
G_2({\cal E}) &=&
\max_{\mathcal{U}_A, \mathcal{U}_B} 
 \min_{\tilde{{\cal E}} \in {\cal P}_{ba}(\omega^\prime)}
 [ H_{\tilde{{\cal E}}}(X|E) - H_{\tilde{{\cal E}}}(X|Y) ]  \\
\label{eq:theorem-bb84-2}
&=&  \min_{\tilde{{\cal E}} \in {\cal P}_{ba}(\omega^*)}
 [ H_{\tilde{{\cal E}}}(X|E) - H_{\tilde{{\cal E}}}(X|Y) ] \\
\label{eq:theorem-bb84-3}
&=& 1 - h\left( \frac{1 + d_\san{z}}{2} \right) - h\left(\frac{1 + d_\san{x}}{2} \right).
\end{eqnarray}
Furthermore, the maximum is achieved without any compensation, i.e.,
\begin{eqnarray*}
\min_{\tilde{{\cal E}} \in {\cal P}_{ba}(\omega)}
 [H_{\tilde{{\cal E}}}(X|E) - H_{\tilde{{\cal E}}}(X|Y)  ]
 = \min_{\tilde{{\cal E}} \in {\cal P}_{ba}(\omega^*)}
 [H_{\tilde{{\cal E}}}(X|E) - H_{\tilde{{\cal E}}}(X|Y) ]
\end{eqnarray*}
if and only if the vectors $\ket{\tilde{U}_\san{z}}$ and $\ket{V_\san{z}}$
are scalar multiple of each other.
\end{theorem}
The first statement implies that an optimal compensation
procedure is to compensate the channel to a channel such
that the left upper sub-matrix $S^\prime$ of the Stokes parameterization
of the compensated channel is a diagonal matrix.
The latter statement implies that $({\cal U}_A, {\cal U}_B)$
\textchange{achieving} the maximum is not unique.

By using Theorem \ref{theorem:bb84}, we can derive the 
following corollary, which gives the key generation rate optimized
by the one-side compensation.
\begin{corollary}
\label{corollary:bb84}
Suppose that ${\cal E}$ is a unital channel.
\textchange{Let}
\begin{eqnarray*}
\textchange{ Q_B^* = \left[ \begin{array}{c}
\bra{Q_{B,\san{z}}^*} \\
\bra{Q_{B,\san{x}}^*}
\end{array} \right] }
\end{eqnarray*}
\textchange{ be a rotation matrix such that
$\bra{Q_{B,\san{z}}^*}$ is a scalar multiple of
$(R_\san{zz}, R_\san{xz})$\footnote{\textchange{Note that
$\bra{Q_{B,\san{x}}}$ is uniquely determined from
$\bra{Q_{B,\san{z}}}$ because they constitute a rotation matrix.}},
\textchangesecond{and let ${\cal U}_B^*$ be the unitary channel
corresponding to $O_B^*$.}
Then, the compensated channel
${\cal U}_B^* \circ {\cal E} =: {\cal E}^* = (\omega^*,\tau^*)$ satisfies}
\begin{eqnarray}
G_1({\cal E}) 
&=& \textchange{ \min_{\tilde{{\cal E}} \in {\cal P}_{ba}(\omega^*)}
 [ H_{\tilde{{\cal E}}}(X|E) - H_{\tilde{{\cal E}}}(X|Y) ] } \\
&=& G_2({\cal E}).
\end{eqnarray}
\textchangesecond{\qed}
\end{corollary}
Note that Corollary \ref{corollary:bb84} follows from the latter
statement of Theorem \ref{theorem:bb84}.

Surprisingly, we do not lose any optimality even if
we only allow Bob to compensate his axis (one-side compensation).
This fact is useful to \textchange{simplify} the implementation of the 
optimal compensation procedure.

Since $H_{{\cal E}}(X|Y) = h((1+ R_\san{zz})/2)$
for any unital channel and $R_\san{zz} =
\braket{V_\san{z}}{\tilde{U}_\san{z}}$, 
we find that
\textchange{an} optimal one-side compensation procedure is to
compensate the channel so that Bob can detect
Alice's transmitted state most reliably, i.e.,
$H_{{\cal E}^\prime}(X|Y)$ is minimized.
Note that the fact that $\ket{\tilde{U}_\san{z}}$ and $\ket{V_\san{z}^\prime}$
is scalar multiple of each other does not 
necessarily mean that the left upper sub-matrix $S^\prime$ 
of the Stokes parameterization of the compensated channel 
is a diagonal matrix.

\noindent{\em Proof of Theorem \ref{theorem:bb84})~}
By using Lemma \ref{lemma:unital-formula},
we have the equality between Eqs.~(\ref{eq:theorem-bb84-2})
and (\ref{eq:theorem-bb84-3}). 
\textchange{Since Eq.~(\ref{eq:theorem-bb84-1}) 
is obviously larger than or equals to
Eq.~(\ref{eq:theorem-bb84-2}),}
it \textchangethird{suffices} to show that
Eq.~(\ref{eq:theorem-bb84-1}) is smaller than or 
equals to Eq.~(\ref{eq:theorem-bb84-3}).
For any fixed ${\cal U}_A$ and ${\cal U}_B$,
by using Lemma \ref{lemma:unital-formula}
again, Eq.~(\ref{eq:theorem-bb84-1}) can be rewritten
as 
\begin{eqnarray*}
1 - h\left(\frac{1 + d_\san{z}}{2} \right)
 - h\left( \frac{1 + d_\san{x}}{2} \right) 
 + h\left(\frac{1 + \| \ket{\tilde{U}^\prime_\san{z}} \| }{2}
 \right)
 - h\left(\frac{1 + \braket{V^\prime_\san{z}}{\tilde{U}^\prime_\san{z}}}{2} \right).
\end{eqnarray*}
From the form of $h(\cdot)$, Cauchy's
inequality
$|\braket{V^\prime_\san{z}}{\tilde{U}^\prime_\san{z}}| \le \| \ket{\tilde{U}^\prime_\san{z}} \|$ implies that Eq.~(\ref{eq:theorem-bb84-1})
is smaller than or equals to Eq.~(\ref{eq:theorem-bb84-3}).
The equality holds if and only if the vectors
$\ket{\tilde{U}^\prime_\san{z}}$ and $\ket{V^\prime_\san{z}}$ are
scalar multiple of each other, which is exactly 
the second statement of the theorem. 
\qed

Next, we consider the case in which Alice and Bob use the
conventional channel estimation. The following theorem
states that the optimized key generation rate with 
the accurate channel estimation coincides with that
with the conventional channel estimation if we use
the two-side compensation.
The following theorem also gives the
necessary and sufficient condition such that the
optimized key generation rates with the accurate
channel estimation and the conventional channel estimation
coincide when we use the one-side compensation.
\begin{theorem}
\label{theorem:bb84-conventional}
Suppose that ${\cal E}$ is a unital channel. Then, we have
\begin{eqnarray*}
G_2({\cal E}) = \tilde{G}_2({\cal E}),
\end{eqnarray*}
\textchangesecond{where $\tilde{G}_2({\cal E})$ is achieved
by $Q_A^*$ and $Q_B^*$ specified in Theorem \ref{theorem:bb84}.}
Furthermore, we have
\begin{eqnarray*}
G_1({\cal E}) = \tilde{G}_1({\cal E})
\end{eqnarray*}
if and only if
$\ket{\tilde{U}_\san{z}}$ and 
$\ket{\tilde{U}_\san{x}}$ are orthogonal \textchange{to} each other.
\textchangesecond{If this condition is satisfied, $\tilde{G}_1({\cal E})$
is achieved by $Q_B^*$ such that 
$\bra{Q_{B,\san{z}}^*}$ and $\bra{Q_{B,\san{x}}^*}$ are scalar multiple of
$(R_\san{zz},R_\san{xz})$ and $(R_\san{zx}, R_\san{xx})$ 
respectively.}
\end{theorem}
\begin{corollary}
Suppose that ${\cal E}$ is a unital channel. Then, we have
\begin{eqnarray*}
\tilde{G}_1({\cal E}) = \tilde{G}_2({\cal E})
\end{eqnarray*}
if and only if
$\ket{\tilde{U}_\san{z}}$ and 
$\ket{\tilde{U}_\san{x}}$ are orthogonal \textchange{to} each other.
\textchangesecond{\qed}
\end{corollary}
\noindent{\em Proof of Theorem \ref{theorem:bb84-conventional})~}
Let ${\cal E}^*$ be the channel defined in 
Theorem \ref{theorem:bb84}. Then, we have
\begin{eqnarray*}
G_2({\cal E}) &\ge&
\tilde{G}_2({\cal E}) \\ 
&\ge& \min_{{\cal E} \in {\cal P}_{bc}(\mu^*)} [
H_{\tilde{{\cal E}}}(X|E) - H_{\tilde{{\cal E}}}(X|Y) ] \\
&=& 1 - h\left( \frac{1 + d_\san{z}}{2} \right) 
  - h\left( \frac{1 + d_\san{x}}{2} \right) \\
&=& G_2({\cal E}),
\end{eqnarray*}
which implies the first statement of the theorem.

To prove the ``if'' part of the second statement, assume that
$\ket{\tilde{U}_\san{z}}$ and $\ket{\tilde{U}_\san{x}}$ 
are orthogonal \textchange{to} each other. 
\textchangesecond{Then, we can take a
rotation matrix $Q_B^*$ so that 
$\bra{Q_{B,\san{z}}^*}$ and $\bra{Q_{B,\san{x}}^*}$ are scalar multiple of
$(R_\san{zz},R_\san{xz})$ and $(R_\san{zx}, R_\san{xx})$ 
respectively, and we have
$S^\prime = Q_B^* S = \mbox{diag}[d_\san{z}, d_\san{x}]$.}
Then, we have
\begin{eqnarray*}
G_1({\cal E}) = \tilde{G}_1({\cal E}) \ge 1 - h\left( \frac{1 + d_\san{z}}{2} \right) 
  - h\left( \frac{1 + d_\san{x}}{2} \right) = G_1({\cal E}).
\end{eqnarray*}

Next, we show the ``only if'' part. Suppose that
$\ket{\tilde{U}_\san{z}}$ and $\ket{\tilde{U}_\san{x}}$ are
not orthogonal \textchange{to} each other. Then, for an arbitrarily fixed 
${\cal U}_B$, either $\braket{V_\san{z}^\prime}{\tilde{U}_\san{x}} \neq 0$
or $\braket{V_\san{x}}{\tilde{U}_\san{z}} \neq 0$ holds. Then, we have
\begin{eqnarray}
\tilde{G}_1({\cal E}) &=&
1 - h\left( \frac{1 + \braket{V_\san{z}}{\tilde{U}_\san{z}}}{2} \right)
 - h\left( \frac{1 + \braket{V_\san{x}}{\tilde{U}_\san{x}}}{2} \right) 
 \nonumber \\
&<& 1 - h\left( \frac{1 + \| \ket{\tilde{U}_\san{z}} \| }{2} \right) 
  - h\left( \frac{1 + \| \ket{\tilde{U}_\san{x}} \| }{2} \right) 
  \nonumber \\
&=& 1 - h\left( \frac{1 + \sqrt{ d_\san{z}^2 U_\san{zz}^2 
   + d_\san{x}^2 U_\san{zx}^2 }}{2} \right) 
  - h\left( \frac{1 + \sqrt{ d_\san{z}^2 U_\san{xz}^2 
   + d_\san{x}^2 U_\san{xx}^2 }}{2} \right) 
  \nonumber \\
&\textchange{\le}& 
\textchange{ 1 - U_\san{zz}^2 h\left(\frac{1 + \sqrt{d_\san{z}^2}}{2} 
  \right) - U_\san{zx}^2 h\left(\frac{1 + \sqrt{d_\san{x}^2}}{2} \right) } 
   \nonumber \\
 &&~~~~~~~~~~~~~
 \textchange{ - U_\san{xz}^2 h\left(\frac{1 + \sqrt{d_\san{z}^2}}{2}\right)
  - U_\san{xx}^2 h\left(\frac{1 + \sqrt{d_\san{x}^2}}{2}\right) } 
  \label{eq:using-concavity} \\
&=& 1 - (U_\san{zz}^2 + U_\san{xz}^2) 
   h\left( \frac{1 + d_\san{z}}{2} \right) 
  - (U_\san{zx}^2 + U_\san{xx}^2) 
   h\left( \frac{1 + d_\san{x}}{2} \right) \nonumber \\
&=& 1 - h\left( \frac{1 + d_\san{z}}{2} \right) 
  - h\left( \frac{1 + d_\san{x}}{2} \right) \nonumber \\
&=& G_1({\cal E}),
 \nonumber 
\end{eqnarray}
where we used the concavity of the function 
\begin{eqnarray}
h\left( \frac{1 + \sqrt{x}}{2} \right)
\label{eq:concave-binary-entropy-function}
\end{eqnarray}
\textchangesecond{in the inequality of Eq.~(\ref{eq:using-concavity}).}
\textchange{We can show the concavity of 
Eq.~(\ref{eq:concave-binary-entropy-function}) by showing
that the second derivative is non-positive.}
\qed

\subsubsection{Compensation within Any Direction}
\label{subsubsec:any-direction-solution}

In this section, we consider the case in which either Alice or 
Bob are allowed to
compensate their axes within any 
direction \cite{chen:07, franson:95, ma:06b, tao:06, xavier:08, zavriyev:05, trifonov:07}.
For any channel ${\cal E} = (R,t)$, by the singular value
decomposition, we can decompose the matrix $R$ as in
Eq.~(\ref{eq:singular-value-decomposition}).
\textchange{Furthermore, we identify
Alice's compensation ${\cal U}_A$  
and Bob's compensation ${\cal U}_B$ with
the $3\times 3$ rotation matrices $O_A$ and $O_B$ as 
in Section \ref{sec:solution-six-state}.}
\textchangesecond{When we consider the compensation within any direction,
it should be noted that we can estimate
all the parameters in the two-side compensation and only
a part of the parameters in the one-side compensation
(see also Section \ref{subsubsec:problem-bb84-2}).}

The following theorem gives the key generation rate
optimized by the two-side compensation.
\begin{theorem}
\label{theorem:bb84-2}
Suppose that ${\cal E}$ is a unital channel. 
\textchange{ 
Let ${\cal U}_A^*$ and ${\cal U}_B^*$ be unitary channels such that 
the compensated channel 
${\cal U}_B^* \circ {\cal E} \circ {\cal U}_A^* =: {\cal E}^* = (\omega^*, \tau^*)$ is a Pauli channel
and the singular values
$|e_\san{z}^*|$, $|e_\san{x}^*|$, and $|e_\san{y}^*|$
of $R^* = \rom{diag}[e_\san{z}^*, e_\san{x}^*, e_\san{y}^*]$
satisfy }
\begin{eqnarray*}
|e_\san{z}^*| \ge |e_\san{x}^*| \ge |e_\san{y}^*|.
\end{eqnarray*}
Then, we have
\begin{eqnarray}
J_2({\cal E}) &=& \max_{{\cal U}_A, {\cal U}_B}
\min_{\tilde{{\cal E}} \in {\cal P}_{ba}(\omega^\prime)}
[ H_{\tilde{{\cal E}}}(X|E) - H_{\tilde{{\cal E}}}(X|Y)] 
  \label{eq:theorem:bb84-2-1} \\
&=&  \min_{\tilde{{\cal E}} \in {\cal P}_{ba}(\omega^*)}
[ H_{\tilde{{\cal E}}}(X|E) - H_{\tilde{{\cal E}}}(X|Y)] 
  \label{eq:theorem:bb84-2-2} \\
&=& 1 - h\left( \frac{1 + e_\san{z}^*}{2} \right)
 - h\left( \frac{1 + e_\san{x}^*}{2} \right).
  \label{eq:theorem:bb84-2-3}
\end{eqnarray}
\end{theorem}

\noindent{\em Proof of Theorem \ref{theorem:bb84-2})~}
By using Lemma \ref{lemma:unital-formula},
we have the equality between Eqs.~(\ref{eq:theorem:bb84-2-2})
and (\ref{eq:theorem:bb84-2-3}). 
\textchange{Since Eq.~(\ref{eq:theorem:bb84-2-1}) is obviously larger 
than or equals to 
Eq.~(\ref{eq:theorem:bb84-2-2}),}
it \textchangethird{suffices} to show that Eq.~(\ref{eq:theorem:bb84-2-1})
is smaller than or equals to Eq.~(\ref{eq:theorem:bb84-2-3}).

For any fixed ${\cal U}_A$ and ${\cal U}_B$, Theorem \ref{theorem:bb84}
implies 
\begin{eqnarray}
\lefteqn{
\min_{\tilde{{\cal E}} \in {\cal P}_{ba}(\omega^\prime)}
[ H_{\tilde{{\cal E}}}(X|E) - H_{\tilde{{\cal E}}}(X|Y)] } \nonumber \\
 &\le& G_2({\cal E}^\prime) \nonumber \\
&=& 1 - h\left( \frac{1 + d_\san{z}^\prime}{2} \right)
  - h\left( \frac{1 + d_\san{x}^\prime}{2} \right),
\label{eq:theorem:bb84-2-4}
\end{eqnarray}
where $|d_\san{z}^\prime|$ and $|d_\san{x}^\prime|$ are 
the singular values of the left upper $2 \times 2$ sub-matrices
$S^\prime$ of $R^\prime$ of the compensated channel ${\cal E}^\prime$.

Note that the singular values of $R^\prime$ are equal to
those of $R^*$. By using the interlacing inequalities for 
singular values of sub-matrices \cite{thompson:72}, we have
\begin{eqnarray*}
|e_\san{z}^*| \ge \max[ |d_\san{z}^\prime|, |d_\san{x}^\prime|]
\end{eqnarray*}
and 
\begin{eqnarray*}
|e_\san{x}^*| \ge \min[ |d_\san{z}^\prime|, |d_\san{x}^\prime|].
\end{eqnarray*}
These inequalities imply that Eq.~(\ref{eq:theorem:bb84-2-4})
is smaller than or equals to Eq.~(\ref{eq:theorem:bb84-2-3}),
which completes the proof. \qed

The following theorem gives the key generation rate
optimized by the one-side compensation.
\begin{theorem}
\label{theorem:bb84-3}
Suppose ${\cal E}$ be a unital channel.
\textchange{Let}
\begin{eqnarray*}
\textchange{
O_B^* = \left[ \begin{array}{c}
\bra{O_{B,\san{z}}^*} \\
\bra{O_{B,\san{x}}^*} \\
\bra{O_{B,\san{y}}^*}
\end{array} \right] }
\end{eqnarray*}
\textchange{ be a rotation matrix such that $\bra{O_{B,\san{z}}^*}$
and $\bra{O_{B,\san{x}}^*}$ span the same subspace
as that spanned by $(R_\san{zz}, R_\san{xz},R_\san{yz})$
and $(R_\san{zx},R_\san{xx},R_\san{yx})$, and that
$\bra{O_{B,\san{z}}^*}$ is a scalar multiple of
$(R_\san{zz}, R_\san{xz}, R_\san{yz})$\footnote{\textchange{Note
that $\bra{O_{B,\san{y}}}$ is uniquely determined from 
$\bra{O_{B,\san{z}}}$ and $\bra{O_{B,\san{x}}}$ because they
constitute a rotation matrix.}},
\textchangesecond{and let ${\cal U}_B^*$ be the unitary
channel corresponding to $O_B^*$.}
Then, the compensated channel 
${\cal U}_B^* \circ {\cal E}=: {\cal E}^* = (\omega^*,\tau^*)$
satisfies }
\begin{eqnarray}
J_1({\cal E}) 
&=& \max_{{\cal U}_B} \min_{\tilde{{\cal E}} \in {\cal
P}_{ba}(\omega^\prime)} [
H_{\tilde{{\cal E}}}(X|E) - H_{\tilde{{\cal E}}}(X|Y) ] 
   \label{eq:theorem:bb84-3-1} \\
&=& \min_{\tilde{{\cal E}} \in {\cal P}_{ba}(\omega^*)} [
H_{\tilde{{\cal E}}}(X|E) - H_{\tilde{{\cal E}}}(X|Y) ]
   \label{eq:theorem:bb84-3-2} \\
&=& 1 - h\left(\frac{1 + s_1^*}{2} \right) 
  - h\left(\frac{1 + s_2^*}{2} \right),
   \label{eq:theorem:bb84-3-3} 
\end{eqnarray}
\textchange{where $s_1^*$ and $s_2^*$ are the singular values of the
upper left $2 \times 2$ sub-matrix matrix }
\begin{eqnarray*}
S^* =  \left[\begin{array}{cc}
\braket{B_\san{z}^*}{\tilde{A}_\san{z}} &
 \braket{B_\san{z}^*}{\tilde{A}_\san{x}} \\
\braket{B_\san{x}^*}{\tilde{A}_\san{z}} &
 \braket{B_\san{x}^*}{\tilde{A}_\san{x}}
\end{array}\right]
\end{eqnarray*}
\textchange{of $R^* = O_B^* R $
such that $s_1^* \ge s_2^*$.}
\end{theorem}

\noindent{\em Proof of Theorem \ref{theorem:bb84-3})~}
The second statement of Theorem \ref{theorem:bb84} implies that 
the equality between Eqs.~(\ref{eq:theorem:bb84-3-2})
and (\ref{eq:theorem:bb84-3-3}).
\textchange{Since Eq.~(\ref{eq:theorem:bb84-3-1}) is 
obviously larger than or equals to 
Eq.~(\ref{eq:theorem:bb84-3-2}),}
we show that Eq.~(\ref{eq:theorem:bb84-3-1})
is smaller than or equals to Eq.~(\ref{eq:theorem:bb84-3-3}).

\textchangesecond{For arbitrarily fixed $O_B$, let
$s_1^\prime$ and $s_2^\prime$ be the singular values
of the upper left $2 \times 2$ matrix}
\begin{eqnarray*}
\textchangesecond{
S^\prime = \left[\begin{array}{cc}
\braket{B_\san{z}^\prime}{\tilde{A}_\san{z}} &
 \braket{B_\san{z}^\prime}{\tilde{A}_\san{x}} \\
\braket{B_\san{x}^\prime}{\tilde{A}_\san{z}} &
 \braket{B_\san{x}^\prime}{\tilde{A}_\san{x}}
\end{array}\right]
}
\end{eqnarray*}
\textchangesecond{of $R^\prime = O_B R$
such that $s_1^\prime \ge s_2^\prime$. }
Then, by using Corollary \ref{corollary:bb84}, we have
\begin{eqnarray}
\lefteqn{ \min_{\tilde{{\cal E}} \in {\cal P}_{ba}(\omega^\prime)}[
H_{\tilde{{\cal E}}}(X|E) - H_{\tilde{{\cal E}}}(X|Y) ] } \nonumber \\
&\le& G_1({\cal E}^\prime) \nonumber \\
&=& 1 - h\left(\frac{1 + s_1^\prime}{2} \right)
  - h\left(\frac{1 + s_2^\prime}{2} \right).
\label{eq:theorem:bb84-3-4}
\end{eqnarray}

By using the minimax principle for singular
values \cite[Problem 3.6.1]{bhatia:book}, we have
\begin{eqnarray}
s_1^\prime 
&=& \max_{x \in \mathbb{R}^2: \|x\| =1} \| S^\prime x \| \nonumber \\
&=& \max_{\alpha,\beta \in \mathbb{R} \atop \alpha^2 + \beta^2 = 1}
\left\| 
\left[\begin{array}{cc}
\braket{B_\san{z}^\prime}{\tilde{A}_\san{z}} &
 \braket{B_\san{z}^\prime}{\tilde{A}_\san{x}} \\
\braket{B_\san{x}^\prime}{\tilde{A}_\san{z}} &
 \braket{B_\san{x}^\prime}{\tilde{A}_\san{x}}
\end{array}\right]
\left[ \begin{array}{c} \alpha \\ \beta \end{array} \right]
\right\| \nonumber \\
&=& \max_{\alpha,\beta \in \mathbb{R} \atop \alpha^2 + \beta^2 = 1}
\left\|
\left[ \begin{array}{c}
\bra{B_\san{z}^\prime}(\alpha\ket{\tilde{A}_\san{z}} + \beta
 \ket{\tilde{A}_\san{x}}) \\
\bra{B_\san{x}^\prime}(\alpha\ket{\tilde{A}_\san{z}} + \beta
 \ket{\tilde{A}_\san{x}}) 
\end{array} \right] \right\| \nonumber \\
&=& \max_{\alpha,\beta \in \mathbb{R} \atop \alpha^2 + \beta^2 = 1}
\sqrt{ \braket{B_\san{z}^\prime}{\Gamma_{\alpha,\beta}}^2 
  + \braket{B_\san{x}^\prime}{\Gamma_{\alpha,\beta}}^2 } \nonumber \\
&\le& \max_{\alpha,\beta \in \mathbb{R} \atop \alpha^2 + \beta^2 = 1}
\sqrt{ \braket{B_\san{z}^*}{\Gamma_{\alpha,\beta}}^2 
  + \braket{B_\san{x}^*}{\Gamma_{\alpha,\beta}}^2 } 
\label{eq:theorem:bb84-3-5} \\
&=& s_1^*,
\label{eq:theorem:bb84-3-6}
\end{eqnarray}
where we set
$\ket{\Gamma_{\alpha,\beta}} := \alpha \ket{\tilde{A}_\san{z}} +
\beta\ket{\tilde{A}_\san{x}}$, and the equality in
Eq.~(\ref{eq:theorem:bb84-3-5}) 
holds if $\ket{B_\san{z}^\prime}$ and $\ket{B_\san{x}^\prime}$
span the same subspace as that spanned by 
$\ket{\tilde{A}_\san{z}}$ and $\ket{\tilde{A}_\san{x}}$.
By using the minimax principle for singular values in a similar
manner, we also have
\begin{eqnarray}
s_2^\prime  = \min_{x \in \mathbb{R}^2: \|x\| = 1} 
\| S^\prime x\|
\le s_2^*.
\label{eq:theorem:bb84-3-7}
\end{eqnarray}

Combining Eqs.~(\ref{eq:theorem:bb84-3-4}),
(\ref{eq:theorem:bb84-3-6}), and 
(\ref{eq:theorem:bb84-3-7}), 
we have shown that Eq.~(\ref{eq:theorem:bb84-3-1})
is smaller than or equals to Eq.~(\ref{eq:theorem:bb84-3-3}). 
\qed

\begin{remark}
The equality 
\begin{eqnarray*}
J_1({\cal E}) = J_2({\cal E})
\end{eqnarray*}
does not hold in general. For example, $J_1({\cal E}) \neq J_2({\cal
 E})$ if $R = \rom{diag}[e_\san{z}, e_\san{x}, e_\san{y}]$ and
$|e_\san{z}| < |e_\san{x}| < |e_\san{y}|$.
\end{remark}

Next, we consider the case in which Alice and Bob use the conventional
channel estimation.
The following theorem states that the optimized key generation rate
with the accurate channel estimation coincides with that
with the conventional channel estimation if we use the
two-side compensation.
The following theorem  also gives the 
necessary and sufficient condition such that the optimized
key generation rates with the accurate channel estimation and
the conventional channel estimation coincide.
\begin{theorem}
\label{theorem:bb84-4}
Suppose that ${\cal E}$ is a unital channel. Then, we have
\begin{eqnarray*}
J_2({\cal E}) = \tilde{J}_2({\cal E}),
\end{eqnarray*}
\textchangesecond{where $\tilde{J}_2({\cal E})$
is achieved by $O_A^*$ and $O_B^*$ specified in
Theorem \ref{theorem:bb84-2}.}
Furthermore, we have
\begin{eqnarray*}
J_1({\cal E}) = \tilde{J}_1({\cal E})
\end{eqnarray*}
if and only if $\ket{\tilde{A}_\san{z}}$ and $\ket{\tilde{A}_\san{x}}$
are orthogonal \textchange{to} each other.
\textchangesecond{If this condition is satisfied, then $\tilde{J}_1({\cal E})$
is achieved by $O_B^*$ such that
$\bra{O_{B,\san{z}}^*}$ and $\bra{O_{B,\san{x}}^*}$
are scalar multiple of 
$(R_\san{zz}, R_\san{xz}, R_\san{yz})$
and $(R_\san{zx}, R_\san{xx}, R_\san{yx})$ 
respectively.}
\end{theorem}
\begin{corollary}
Suppose that ${\cal E}$ is a unital channel.
Then, we have
\begin{eqnarray*}
\tilde{J}_1({\cal E}) < \tilde{J}_2({\cal E})
\end{eqnarray*}
if $\ket{\tilde{A}_\san{z}}$ and $\ket{\tilde{A}_\san{x}}$
are not orthogonal \textchange{to} each other.
\textchangesecond{\qed}
\end{corollary}
\noindent{\em Proof of Theorem \ref{theorem:bb84-4})~}
This theorem can be proved almost in a similar manner \textchangethird{to}
Theorem \ref{theorem:bb84-conventional}.
Therefore, we omit the proof.
\qed

\section{Conclusion}
\label{sec:conclusion}

\textchangesecond{In this paper, we investigated the axis compensation
in the QKD protocols \textchangethird{in} various settings.
\textchangethird{We clarified optimal compensation procedures
over unital channels
for one-side compensation with the accurate channel estimation
and for two-side compensation with both estimation, while we could
not identify an optimal compensation procedure for one-side compensation
with the conventional channel estimation.
Although our proposed compensation procedures
are optimal for unital channels, it is not clear whether
those compensation procedures are optimal or not for
general channels.}
We also clarified that the optimized key generation
rates with the conventional channel estimation are
strictly smaller than the optimized key generation
rates with the accurate channel estimation for the one-side compensation.}
Our \textchange{results} imply that
we should use the accurate channel estimation when
we employ the one-side compensation.
On the other hand, we do not have to use
the accurate channel estimation when we employ 
the two-side compensation.

Although we clarified the optimal compensation
procedures for the standard postprocessing,
it is an important future research agenda to clarify the
optimal compensation procedures when we employ more 
complicated postprocessing (e.g.~the postprocessing
with the noisy preprocessing \cite{renner:05, kraus:05} 
or the two-way classical communication
\cite{gottesman:03, watanabe:07}).

\section*{Acknowledgment}
The authors would like to thank Dr.~Toyohiro Tsurumaru
for bringing the axis
compensation problem to our attention.
\textchange{The first author would like to thank Prof. Yasutada Oohama for
his support.}
This research is partly supported by the Japan
Society of Promotion of Science under 
Grants-in-Aid No.~00197137.



\end{document}